# Real-Time Elderly Healthcare Monitoring Expert System Using Wireless Sensor Network

[a]Ibrahim Almarashdeh, [a]Mutasem K. Alsmadi, [a]Tamer Farag, [b]Abdullah S. Albahussain, [a]Usama A Badawi, [a]Njoud Altuwaijri, [a]Hala Almaimoni, [a]Fatima Asiry, [a]Shahad Alowaid, [a]Muneerah Alshabanah, [a]Daniah Alrajhi, [c]Amirah Al Fraihet, [d]Ghaith Jaradat

[a]College of Applied Studies and Community Service, Imam Abdurrahman Bin Faisal University, Al-Dammam, Saudi Arabia

[b]Family Medicine Resident, Family and Community Medicine Department, Imam Abdurrahman Bin Faisal University, Al-Dammam, Saudi Arabia

[c]Faculty of Pharmacy, Al-Zaytoonah University of Jordan, Amman, Jordan.

[d]Department of Computer Science, Faculty of Information Technology, Jerash University, 26150-311 Jerash, Jordan.

Email: mksalsmadi@gmail.com, ibramars@gmail.com

***Abstract:*** Elderly chronic diseases are the main cause of death in the world, accounting 60% of all death. Because elderly with chronic diseases at the early stages has no observed symptoms, and then symptoms starts to appear, it is critical to observe the symptoms as early as possible to avoid any complication. This paper presents an expert system for an Elderly Health Care (EHC) at elderly home tailored for the specific needs of Elderly. The proposed EHC aims to develop an integrated and multidisciplinary method to employ communication technologies and information for covering real health needs of elderly people, mainly of people at high risk due to social and geographic isolation in addition to specific chronic diseases. The proposed EHC provides personalized intervention plans covering chronic diseases such as (body temperature (BT), blood pressure (BP), and Heart beat rate (HR)). The processes and architecture of the proposed EHC are based on the server side and three main clients, one for the elderly and another two for the nurse and the physician's whom take care of them. The proposed EHC model is discussed for proving the usefulness and effectiveness of the expert system.

Keywords: Diagnosis system; smart help care; health technology; chronic diseases; Elderly Health Care (EHC).

## 1. Introduction

Population around the world is ageing [1, 2], this is due to number of factors such as decreased fertility rates and increased expectancy of life, mainly linked to the increased birth control and movements of migration. This trend is more evident and continuously growing in developed countries, because the number of elderly people is already large [1, 3-7]. Aging has impact on the social and economic foundations of communities, thus; governments need to employ more funding for taking care of the elderly people and fewer hands are available cover their needs [4].

For the health care systems, this is considered as a big challenge [8], these health care systems need to deal with the health issues of the elderly population [3, 5-7, 9-16]. In order to develop effective health systems, they must be able to deal with the so-called "four giants" of geriatrics (Isaacs) as: instability, immobility, intellectual impairment and incontinence. Moreover, chronic diseases are widely spread in elderly, such as Alzheimer's, chronic respiratory diseases, cardiovascular diseases and diabetes [4]. Thus, mobile application widely used in our daily life. On of reasons of using mobile application in education, government services and health care is the usability, ease of use and usefulness of using the application at anywhere and anytime [17-19]. Using mobile application to track elderly people is one of the benefits of such services. On other hand, accessing information and monitoring the up-to-date issues is the best benefit of mobile application [20, 21]. Using mobile application in health care would help doctors follow up with patient and track their status without the hassle of traveling to hospital.

According to the World Health Organization [22], elderly chronic diseases are the main cause of death in the world, accounting 60% of all death. In Europe this proportion is higher, where these diseases are estimated to account for 77 % of the disease burden and an 86 % of the total deaths in the area [23]. Thence, it is essential to develop effective health systems that are able to keep the chronic diseases under control.

The Active Ageing, which is the process of improving opportunities for elderly's health, security and participation which aims to improve quality of life as individuals age, is portion of the process of handling these chronic diseases, to make effort to keep the elderly healthy and participating in the social live [4]. Recently; artificial intelligence algorithms were widely used for solving very complicated problems, such as patterns recognition and information retrieval [24-33], image segmentations [34-40] and river flow forecasting [41], Many researchers employed communication technologies and information [6, 42], and artificial intelligence algorithms [43-46] for covering real health needs of patients.

The proposed EHC aims to develop an integrated and multidisciplinary method to employ communication technologies and information for covering real health needs of elderly people. The system will help the elderly to prevent or keep under control their chronic diseases as well as make the physician's aware of the chronic diseases of the elderly and help them when necessary. The work will be useful for the professionals giving them ways for treatments' prescription (medical treatments) for the elderly patients and for a following up of the patients. Reduce the chance of life taken away. In conclusion, this work also profits the health control products, which are incorporated within the systems to enable treatments' follow up.

## 2. Related works

A pervasive monitoring system to transfer the patients physical signs to real time remote medical applications was proposed in [47]. The system had two components: data acquisition section and data transmission section. The health parameters (ECG, blood pressure, heart rate, SpO2, blood fat, pulse rate and blood glucose) in addition to other indicator (location of the patient) are planned to be sampled continuously at different rates. Four modes of data transmission are used medical analysis needs, taking risk, computing resources into consideration and demands for communication. Finally, the authors implemented a sample prototype to provide a system overview. Figure 1 shows the monitoring system architecture.

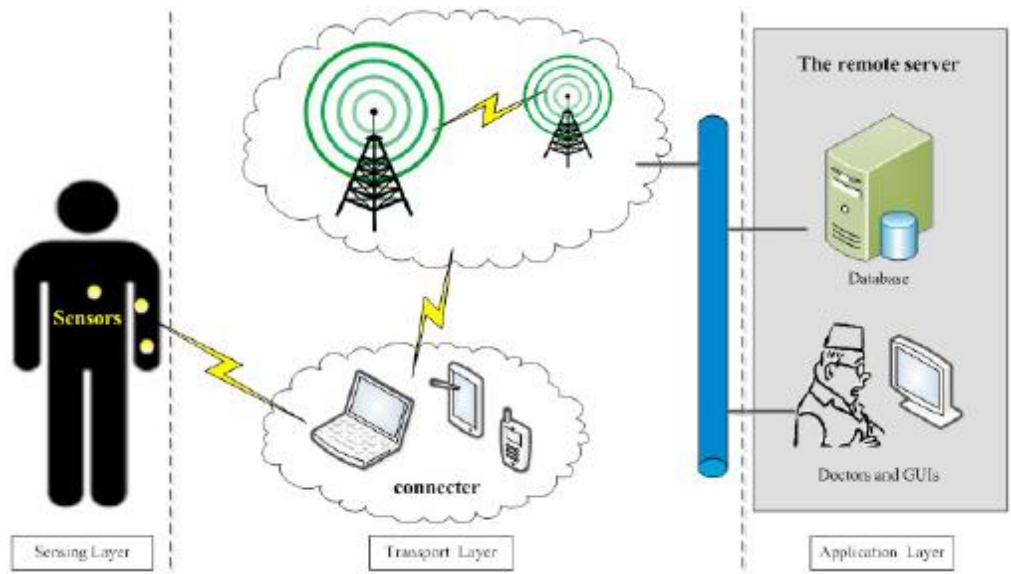

Figure 1: shows the monitoring system architecture [47].

Ronnie et al., in [48] offered a framework for u-Healthcare system using Radio Frequency Identification (RFID) and Wireless Medical Sensor Networks (WMSN). The system of monitors the medical status of the patient by using body sensor with the RFID and transmits the data by the wireless to the nearest local workstation (WMSN gateway) then transmits it to the central server. Proper medical services are administered locally in the workstation depending on queries with the central database that has the patient's information. In their work, patients are alerted when there is an emergency case and receive alert message with their smart phones. The medical staff on the workstation will also receive a messages showing that the patient's health needs attention. Based on the patient's health status a medical service will be applied or prescribed to the patient. Figure 2 shows the U-healthcare system design using WMSN and RFID.

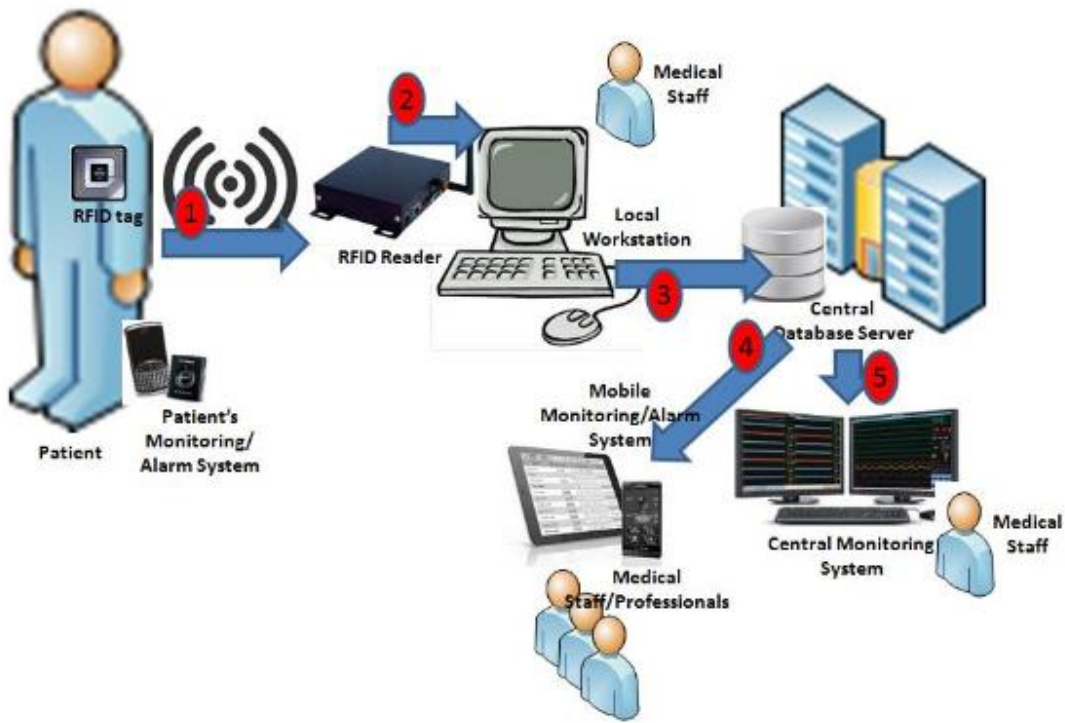

Figure 2: The U-healthcare system design using WMSN and RFID [48].

Al-Aubidy et al., in [42] designed and implemented a real-time healthcare monitoring system. The system scans, calculates, monitors and communicates with medical center's using a group of sensors connected to microcontroller which has a tool for wireless communication to transmit the real-time health information from the patient to the medical center which helps to detect any abnormal medical condition.

Elderly health monitoring system using smart home gateway was proposed in [9]. The system has provided continuous and long-term monitoring for the elderly. In consideration of the mass data generated in the monitoring process, an ECG compression algorithm was built and tested. The results show that it is possible to apply the algorithm to the real-time monitoring system. Figure 3 shows the Architecture health monitoring system.

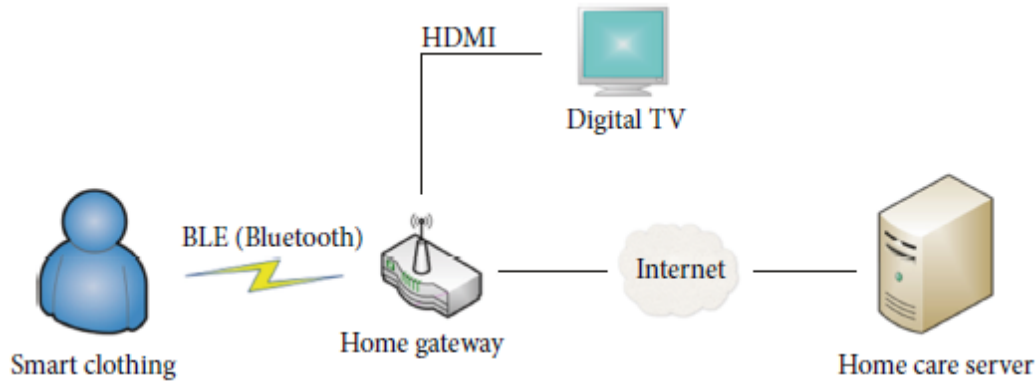

Figure 3: Shows the Architecture health monitoring system [9].

Gogate and Bakal in [49] presented a healthcare model using WSN to remove limitations of wired system and to utilize wireless technology efficiently. The system continuously screens the patient's body temperature and heart rate and when detecting any abnormal parameter, sending email alert or sms to the physician, nurse and close relative of the patient for emergency handling.

3. The Proposed EHC Healthcare Expert System Design

The EHC expert system can be divided into three sections; the smart bracelets with embed medical sensors and it uses wireless technology to communicate with DB server, and a LAN that connect the medical center unit which is consist of DB server and nurses' computers, the third section is the physicians' PDAs connected to the DB server through internet infrastructure communications, that is illustrated in figure 4.

Each patient smart bracelet is considered as a node of a Wireless Sensors Networks (WSN) with Gateway node at DB server. Each nurse computer connected to the DB server node through LAN connection. The DB server node receives the health indicator measurements from the bracelets, it stores the measurements, and send these measurements to nurses' computers. In the case of critical measurements recorded, flags sent by the DB server to the physician's PDAs.

The bracelet unit is considered as wireless sensor node, and it is equipment with microcontrollers and embed health indicator sensors and as shown in Fig 4. The EHC designed with the use of low power and micro medical indicator sensors. They are four medical indicator sensors to measure the body temperature (BT), blood pressure (BP), and Heart beat rate (HR).

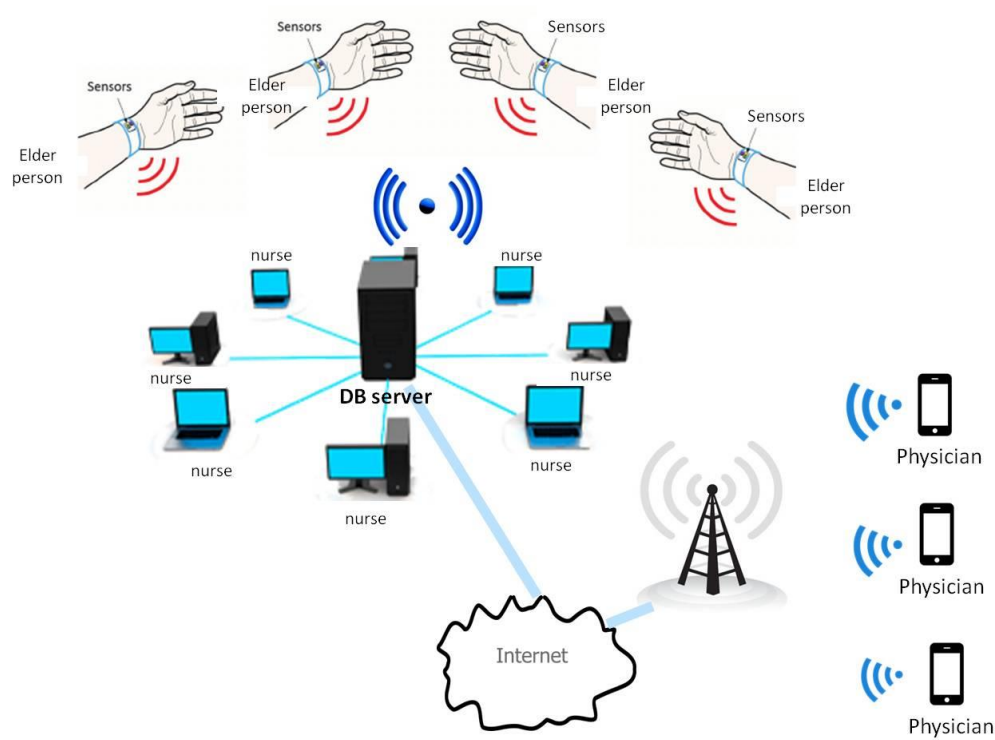

Figure 4: the proposed healthcare expert system architecture design.

The physician can have checked the elder health syndrome by using his PDA or his PC through internet connection to the DB server.

The DB server employs a healthcare software system to analyze the received sensed indicator, that software is equipment with an expert system for health care domain, it functionality can be described as follows:

- Receiving the indicator values from the bracelet sensors.
- Analyze the measurements to define any abnormal or vital health condition.
- Illustrate the received data and analyzed information of the elders' health condition to the nurses' computers
- Determine if any necessary condition must be reported to the physicians' PDAs using the expert system.

- Get any recommendation by the physician and added to the DB and also update the expert system knowledge base if needed.

The database has a GUI built using PHP and MySQL, as illustrated in Fig 5. The database system contains table of the elder to show the personal information of each elder with the ability of find a specific person by his name and/or his ID vaue. When a elder is selected, his/her information can be modified, and if there is any abnormal measurements indicators, that will appear and stored as a note. Besides that, it provides the date and time of last update of his biomedical data. Medical conditions query shows the data for the specified patient, such as all normal and abnormal readings stored in the database.

Figure 5: The database has a GUI built using PHP and MySQL.

The medical staff on their workstations will receive an alert message indicating that the patient status if it needs attention. As well as, Physician will be alerted in case of emergency by sending a message to their PDAs. A medical service depending on the status of the patient will be prescribed or applied to the patient.

When the physician login to his application interface, he will be directly being able to see the data received by the system for every individual patient along with their current and previous medical reports in the database system with their statistical charts. The physician can view the patient medical history on the interface application on his PDA. So, based on the previous medical records and current measurements the physician can then prescribe the medication in acknowledgement to the patients' data. This prescription would be immediately visible to patient along with the physician registration number and other valid credential details, as shown in Fig. 5.

## 4. System discussion

By getting the correct information about the elderly health at the right time, the proposed EHC system will help the nurses and physician through following advantages:

1. Assists the physician in tracking and monitoring the elderly health from a distance.
2. Gives alerts about changes in the elderly health and reduce the complications.
3. Monitors the condition of the elderly periodically.
4. Contribute to saving the lives of elderly.
5. Help the nurse to ask for help through the application by the alert that appears to the physician's when any sudden change in (pressure - heartbeat - body temperature - sugar) happens.

The prototype of the proposed EHC system has been verified and tested to be sure that all of the software and hardware components are working accurately. The smart bracelet with embedded medical sensors has a wireless technology to connect the elderly side with DB server, as well as the physicians' PDAs are connected to the DB server through internet infrastructure communications. The programmed database contains the patient's details, medical conditions, medical history, medications and patient's health status evaluation. Moreover; the database contains the real data received from smart bracelets including the BT, BP and HR. the user interface was designed to help the nurses and physicians to monitor and track the elderly's real-time health parameters. The main limitation of this proposed EHC is the smart bracelet with set of integrated medical sensors, due to the limited fund for this project the smart bracelet in this stage is not fully functionally work. Therefore; in this stage we used a nurse to replace the smart bracelet roles. The future work is to design and implement the smart bracelet with set of integrated medical sensors for elderly healthcare monitoring system to transfer real-time medical information between the elderly and the medical center.

## 5. Conclusion

This proposed EHC expert system can be a key element of intrinsically complicated E-healthcare systems that are being planned and developed to cope with increasingly stressed elderly healthcare infrastructure. A prototype model is proposed for monitoring elderly patients by the smart bracelets with a set of integrated medical sensors to provide the nursing staff & medical experts with real-time health data. A notification system is proposed to inform duty medical staff and physician's and relatives as well in case of any emergency. The medical staff on the workstation will receive an alert message indicating that the elderly patient status needs attention. As well as, Physician can be alerted in case of emergency through the medical staff computer by sending messages to their Smartphones. A medical service depending on the status of the patient will be prescribed or applied to the patient. This proposed expert system can be enhanced by acquiring other health parameters from the elderly patient's.

**References**

[1]   F. Touati and R. Tabish, "U-Healthcare System: State-of-the-Art Review and Challenges," *Journal of Medical Systems,* vol. 37, p. 9949, May 03 2013.
[2]   K. Nisar, A. A. A. Ibrahim, L. Wu, A. Adamov, and M. J. Deen, "Smart home for elderly living using Wireless Sensor Networks and an Android application," in *Application of Information and Communication Technologies (AICT), 2016 IEEE 10th International Conference on*, 2016, pp. 1-8.
[3]   R. J. Conejar and H. Kim, "Proposed Architecture for U-Healthcare Systems," *Int. J. Softw. Eng. Appl,* vol. 9, pp. 213-218, 2015.
[4]   I. Martí Ruiz, J. P. Lázaro Ramos, and A. Aracil Ramón, "Remote health care system for elderly people with chronic diseases," 2013.
[5]   Y. Zhang, H. Liu, X. Su, P. Jiang, and D. Wei, "Remote mobile health monitoring system based on smart phone and browser/server structure," *Journal of Healthcare Engineering,* vol. 6, pp. 717-738, 2015.
[6]   J. Chauhan and S. Bojewar, "Sensor networks based healthcare monitoring system," in *Inventive Computation Technologies (ICICT), International Conference on*, 2016, pp. 1-6.
[7]   M. C. Selvi, T. D. Rajeeve, A. J. P. Antony, and T. Prathiba, "Wireless sensor based healthcare monitoring system using cloud," in *2017 International Conference on Inventive Systems and Control (ICISC)*, 2017, pp. 1-6.

[8] M. Rasmi, M. B. Alazzam, M. K. Alsmadi, I. A. Almarashdeh, R. A. Alkhasawneh, and S. Alsmadi, "Healthcare professionals' acceptance Electronic Health Records system: Critical literature review (Jordan case study)," *International Journal of Healthcare Management,* pp. 1-13, 2018.

[9] K. Guan, M. Shao, and S. Wu, "A Remote Health Monitoring System for the Elderly Based on Smart Home Gateway," *Journal of Healthcare Engineering,* vol. 2017, 2017.

[10] U. Gogate, R. Panat, O. Ninawe, and S. Soma, "Health Care System Based on Wireless Sensor Networks," *International Journal of Innovative Research in Computer and Communication Engineering,* vol. 4, pp. 3311-3315, 2016.

[11] A. Yadav, V. Kaundal, A. Sharma, P. Sharma, D. Kumar, and P. Badoni, "Wireless Sensor Network Based Patient Health Monitoring and Tracking System," in *Proceeding of International Conference on Intelligent Communication, Control and Devices : ICICCD 2016*, R. Singh and S. Choudhury, Eds., ed Singapore: Springer Singapore, 2017, pp. 903-917.

[12] J. Ma, N. Y. Yen, R. Huang, and X. Zhao, "W2T Framework Based U-Pillbox System Towards U-Healthcare for the Elderly," in *Wisdom Web of Things*, N. Zhong, J. Ma, J. Liu, R. Huang, and X. Tao, Eds., ed Cham: Springer International Publishing, 2016, pp. 209-236.

[13] M. Al Hemairy, M. Serhani, S. Amin, and M. Alahmad, "A Comprehensive Framework for Elderly Healthcare Monitoring in Smart Environment," in *Technology for Smart Futures*, ed: Springer, 2018, pp. 113-140.

[14] R. Gravina, P. Alinia, H. Ghasemzadeh, and G. Fortino, "Multi-sensor fusion in body sensor networks: State-of-the-art and research challenges," *Information Fusion,* vol. 35, pp. 68-80, 2017.

[15] H. Mshali, T. Lemlouma, and D. Magoni, "Adaptive monitoring system for e-health smart homes," *Pervasive and Mobile Computing,* vol. 43, pp. 1-19, 2018.

[16] P. Nedungadi, A. Jayakumar, and R. Raman, "Personalized Health Monitoring System for Managing Well-Being in Rural Areas," *Journal of Medical Systems,* vol. 42, p. 22, 2018.

[17] I. Almarashdeh and M. Alsmadi, "Investigating the acceptance of technology in distance learning program," in *Information Science and Communications Technologies (ICISCT), International Conference on*, Tashkent Uzbekistan 2016, pp. 1-5.

[18] I. Almarashdeh and M. Alsmadi, "Heuristic evaluation of mobile government portal services: An experts' review," in *2016 11th International Conference for Internet Technology and Secured Transactions (ICITST)*, 2016, pp. 427-431.

[19] I. Almarashdeh and M. K. Alsmadi, "How to make them use it? Citizens acceptance of M-government," *Applied Computing and Informatics,* vol. 13, pp. 194-199, 2017/07/01/ 2017.

[20] I. Almarashdeh, A. Althunibat, and N. F. Elias, "Developing a Mobile Portal Prototype for E-government Services," *Journal of Applied Sciences,* vol. 14, pp. 791-797, 2014.

[21] I. Almarashdeh, "Sharing instructors experience of learning management system: A technology perspective of user satisfaction in distance learning course," *Computers in Human Behavior,* vol. 63, pp. 249-255, 2016.

[22] WHO, "(2011), World Health Organization, Chronic diseases. http://www.who.int/topics/chronic_diseases/en/."

[23] WHO-EU, "(2011). Non-communicable diseases in Europe Region. http://www.euro.who.int/en/what-we-do/health- topics/noncommunicable-diseases."

[24]     A. M. Al Smadi, M. K. Alsmadi, H. Al Bazar, S. Alrashed, and B. S. Al Smadi, "Accessing Social Network Sites Using Work Smartphone for Face Recognition and Authentication," *Research Journal of Applied Sciences, Engineering and Technology,* vol. 11, pp. 56-62, 2015.

[25]     M. Alsmadi, "Facial recognition under expression variations," *Int. Arab J. Inf. Technol.,* vol. 13, pp. 133-141, 2016.

[26]     M. Alsmadi and K. Omar, *Fish Classification: Fish Classification Using Memetic Algorithms with Back Propagation Classifier*: LAP LAMBERT Academic Publishing, 2012.

[27]     M. Alsmadi, K. Omar, S. Noah, and I. Almarashdeh, "A hybrid memetic algorithm with back-propagation classifier for fish classification based on robust features extraction from PLGF and shape measurements," *Information Technology Journal,* vol. 10, pp. 944-954, 2011.

[28]     M. Alsmadi, K. B. Omar, S. A. Noah, and I. Almarashdeh, "Fish Recognition Based on Robust Features Extraction from Size and Shape Measurements Using Neural Network " *Journal of Computer Science,* vol. 6, pp. 1088-1094, 2010.

[29]     M. K. Alsmadi, "An efficient similarity measure for content based image retrieval using memetic algorithm," *Egyptian Journal of Basic and Applied Sciences*.

[30]     M. K. Alsmadi, "Query-sensitive similarity measure for content-based image retrieval using meta-heuristic algorithm," *Journal of King Saud University - Computer and Information Sciences*.

[31]     M. K. Alsmadi, A. Y. Hamed, U. A. Badawi, I. Almarashdeh, A. Salah, T. H. Farag, W. Hassan, G. Jaradat, Y. M. Alomari, and H. M. Alsmadi, "FACE IMAGE RECOGNITION BASED ON PARTIAL FACE MATCHING USING GENETIC ALGORITHM," *SUST Journal of Engineering and Computer Sciences (JECS),* vol. 18, pp. 51-61, 2017.

[32]     M. K. Alsmadi, K. B. Omar, S. A. Noah, and I. Almarashdeh, "Fish recognition based on robust features extraction from color texture measurements using back-propagation classifier," *Journal of Theoritical and Applied Information Technology,* vol. 18, 2010.

[33]     U. A. Badawi and M. K. Alsmadi, "A GENERAL FISH CLASSIFICATION METHODOLOGY USING META-HEURISTIC ALGORITHM WITH BACK PROPAGATION CLASSIFIER," *Journal of Theoretical & Applied Information Technology,* vol. 66, 2014.

[34]     T. H. Farag, W. A. Hassan, H. A. Ayad, A. S. AlBahussain, U. A. Badawi, and M. K. Alsmadi, "Extended Absolute Fuzzy Connectedness Segmentation Algorithm Utilizing Region and Boundary-Based Information," *Arabian Journal for Science and Engineering,* pp. 1-11, 2017.

[35]     Z. Thalji and M. Alsmadi, "Iris Recognition using robust algorithm for eyelid, eyelash and shadow avoiding," *World Applied Sciences Journal,* vol. 25, pp. 858-865, 2013.

[36]     M. K. Alsmadi, "A hybrid Fuzzy C-Means and Neutrosophic for jaw lesions segmentation," *Ain Shams Engineering Journal*.

[37]     U. A. Badawi and M. K. S. Alsmadi, "A Hybrid Memetic Algorithm (Genetic Algorithm and Great Deluge Local Search) With Back-Propagation Classifier for Fish Recognition " *International Journal of Computer Science Issues,* vol. 10, pp. 348-356, 2013.

[38]     A. M, O. K, and N. S, "Back Propagation Algorithm : The Best Algorithm Among the Multi-layer Perceptron Algorithm," *International Journal of Computer Science and Network Security,* vol. 9, pp. 378-383, 2009.

[39]     M. k. Alsmadi, K. B. Omar, S. A. Noah, and I. Almarashdah, "Performance Comparison of Multi-layer Perceptron (Back Propagation, Delta Rule and Perceptron) algorithms in Neural Networks," in *2009 IEEE International Advance Computing Conference*, 2009, pp. 296-299.

[40]  M. k. Alsmadi, K. B. Omar, and S. A. Noah, "Proposed method to decide the appropriate feature set for fish classification tasks using Artificial Neural Network and Decision Tree," *IJCSNS* vol. 9, pp. 297-301, 2009.

[41]  M. K. Alsmadi, "Forecasting River Flow in the USA Using a Hybrid Metaheuristic Algorithm with Back-Propagation Algorithm," *Scientific Journal of King Faisal University (Basic and Applied Sciences),* vol. 18, pp. 13-24, 2017.

[42]  K. Al-Aubidy, A. Derbas, and A. Al-Mutairi, "Real-time healthcare monitoring system using wireless sensor network," *International Journal of Digital Signals and Smart Systems,* vol. 1, pp. 26-42, 2017.

[43]  M. K. Alsmadi, "A hybrid firefly algorithm with fuzzy-C mean algorithm for MRI brain segmentation," *Am. J. Applied Sei,* vol. 11, pp. 1676-1691, 2014.

[44]  M. K. Alsmadi, "Mri brain segmentation using a hybrid artificial bee colony algorithm with fuzzy-c mean algorithm," *Journal of Applied Sciences,* vol. 15, p. 100, 2015.

[45]  M. K. Alsmadi, "A hybrid Fuzzy C-Means and Neutrosophic for jaw lesions segmentation," *Ain Shams Engineering Journal,* 2017.

[46]  E. Vázquez-Santacruz and M. Gamboa-Zúñiga, "An intelligent device for assistance in caring for the health of patients with motor disabilities," in *2015 11th International Conference on Natural Computation (ICNC)*, 2015, pp. 868-874.

[47]  C. Li, X. Hu, and L. Zhang, "The IoT-based heart disease monitoring system for pervasive healthcare service," *Procedia Computer Science,* vol. 112, pp. 2328-2334, 2017/01/01/ 2017.

[48]  R. D. Caytiles and S. Park, "A study of the design of wireless medical sensor network based u-healthcare system," *International Journal of Bio-Science and Bio-Technology,* vol. 6, pp. 91-96, 2014.

[49]  U. Gogate and J. W. Bakal, "Smart Healthcare Monitoring System based on Wireless Sensor Networks," in *Computing, Analytics and Security Trends (CAST), International Conference on*, 2016, pp. 594 - 599.